# Effective Resistivity in Collisionless Magnetic Reconnection

## Z. W. Ma* and T. Chen

Institute for Fusion Theory and Simulation, Department of Physics, Zhejiang University, Hangzhou 310027, China.

**Abstract**

Magnetic reconnection (MR) in collisionless plasma is often attributed to the off-diagonal electron Reynolds stress, which can give rise to a large induction electric field in the reconnection region. However, in magneto-hydro-dynamics (MHD) simulations of MR, it is difficult to implement the full Reynolds stress, which is kinetic in nature. In this paper, we propose a theoretical model of effective resistivity from the first principle of particle dynamics. The derived theoretical formulation of the effective resistivity is verified by full particle-in-cell (PIC) simulations, and the corresponding physics is discussed.

* Corresponding author: Z. W. Ma (zwma@zju.edu.cn)

# 1 Introduction

MR is an important plasma process that efficiently converts magnetic energy into plasma kinetic and thermal energies (Dungey, 1961; Vasyliunas, 1975) and is believed to play crucial roles in the evolution of the solar corona (Hesse et al., 2005; Hsieh et al., 2009; Kopp & Pneuman, 1976), geomagnetic tail (Bhattacharjee, 2004; Birn & Hesse, 1991; Oieroset et al., 2001), magnetosphere (Deng & Matsumoto, 2001; Goldstein et al., 1986), as well as laboratory fusion plasmas (Furth et al., 2015; Wang & Ma, 2015).

In collisionless plasma, a widely accepted physical mechanism for fast MR (FMR) is an increase of the effect of the off-diagonal (with respect to the ambient magnetic field) electron Reynolds stress in the diffusion region, which gives rise to a large reconnection electric field that strongly accelerates the charged particles in the region (Cai & Lee, 1997; Pritchett, 2001). However, the Reynolds stress is associated with the electron kinetic effects and can therefore not be easily implemented in fluid descriptions of the plasma. In many MHD models, FMR is attributed to anomalous resistivity arising from current-instability driven turbulence in the diffusion region (Malyshkin et al., 2005; Ugai, 1984). However, such an anomalous resistivity often involves artificially given (usually constant) turbulence level or is only current dependent. Speiser (1970) has introduced an effective conductivity for studying collisionless FMR without invoking turbulence. However, the model does not include the details of the particle motion that give rise to the effective conductivity, so that it is not clear how particles are accelerated.

In this paper, we introduce an effective resistivity for considering collisionless FMR. The effective resistivity is obtained by replacing the collision mean-free-time in the traditional collisional drag force with the transit time of electrons in the small diffusion region around the X point of the MR. The transit time is obtained by following the motion of test electrons in the region and, as to be expected, is space and time dependent. Validity of our ad hoc model is confirmed by full PIC simulation.

The rest of this paper is as follows. Section 2 presents our effective resistivity model and its properties. A theoretical argument justifying the effective resistivity is also given. Section 3 presents the corresponding PIC simulation. Section 4 compares the results from the model and the simulation. Section 5 gives a summary of our work.

# 2 Physical Mechanisms and Model Description

Classically, plasma resistivity arises from inter-particle collisions that lead to momentum and energy exchange between the colliding particles. It therefore depends on the collision frequency or the mean free path. In collisionless MR, particles in the small diffusion region around the X point experience strong electric and magnetic forces. A particle is first decelerated, and then accelerated as it enters and leaves the diffusion point due to the bent magnetic fields and the induction electric fields. It thereby exchanges energy with the fields. The interaction can thus lead to a local effective resistivity around the X point in a region of the order of the electron inertial length. The scenario is roughly similar to what occurs in a binary collision, namely the interaction takes place in a very small region around the center of mass or a massive particle, analogous to the X point in MR.

We consider the dynamics of a charge particle along an X line (assumed to be in the z direction, perpendicular to the MR plane) of the diffusion region, where the magnetic field is nearly zero and the induction electric field $E_z$ is strong. The change in the velocity of the particle can be written as Speiser (1970) does

$$\delta v_z = qE_z \delta t / m, \tag{1}$$

where $q$ and $m$ are the particle charge and mass, respectively, and $\delta t$ is the transit time of the particle. Accounting for all the particles in the diffusion region, the corresponding change in the local current density is

$$\delta J_z = nq\delta v_z = nq^2 E_z \delta t / m, \tag{2}$$

where $n$ is the local particle density. Thus, one can define an effective resistivity

$$\eta \equiv E_z / J_z = m / nq^2 \delta t, \tag{3}$$

which is valid only near the X point. One must however still determine the particle density and the transit time.

To model the diffusion region in collisionless MR, we consider a two-dimensional (2D) plane ($x, y$) with the X line lying in the perpendicular, or $z$, direction at ($0, 0$). The vacuum magnetic and electric fields in this region can be approximated by

$$\boldsymbol{B} = B_0 \frac{y\hat{\boldsymbol{x}}}{L_y} + B_{y0} \frac{xy}{L_x}, \tag{4}$$

$$\boldsymbol{E} = E_0 \hat{\boldsymbol{z}}, \tag{5}$$

where $B_0$ and $E_0$ are positive constants, $B_{y0}$ is the asymptotical reconnected magnetic field, $L_x$ and $L_y$ are the local characteristic lengths of $B_y$ and $B_x$ in the $x$ and $y$ directions, respectively. That is, the induction electric field remains uniform in this region, and the magnetic field increases with the distance away from the X line (or X point in the ($x, y$) plane).

We first consider a general case of the motion of a test electron in the diffusion region. Initially, the electron is at $(x_0, y_0)$ and its velocity components are $v_{x0} > 0$, $v_{y0} > 0$, $v_{z0} < 0$. The region considered is $2L_x \times 2L_y$. The fields and other parameters are illustrated in Figure 1. The configuration here differs from that of Speiser's model (1970), where the diffusion region is one-dimensional. It is similar to the model in Moses et al.'s paper in 1993, except that here more details are involved, such that the FMR process can be better understood.

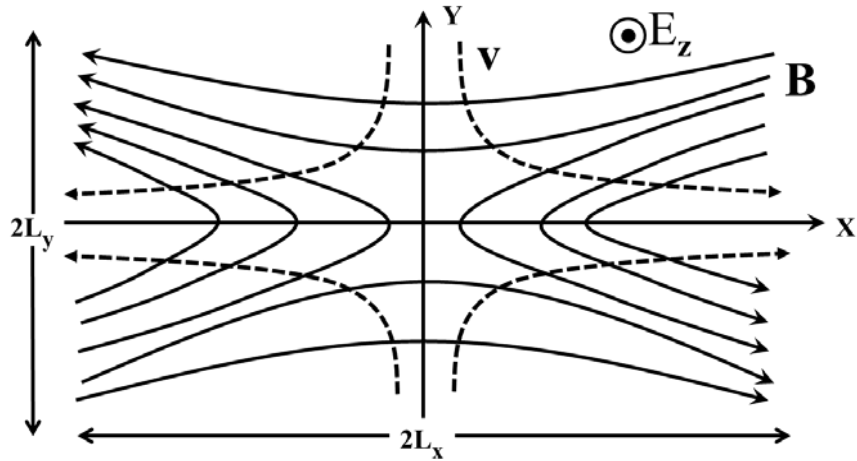

Figure 1  Schematics of magnetic field lines and electron trajectories in the 2D diffusion region. The X line in the $z$ direction is at (0,0).

An electron inside this box will be driven by the Lorentz and electric forces:

$$F_x = -qv_z B_{y0} x / L_x, \tag{6}$$

$$F_y = qv_z B_0 y / L_y, \tag{7}$$

$$F_z = q(E_0 + v_x B_{y0} x / L_x - v_y B_0 y / L_y). \tag{8}$$

The trajectory of the electron is then given by

$$x(t) = x_0 + \int_0^t [v_{x0} - \int_0^{t'} \frac{q}{mL_x} v_z(t'') B_{y0} x(t'') dt''] dt', \qquad (9)$$

$$y(t) = y_0 + \int_0^t [v_{y0} + \int_0^{t'} \frac{q}{mL_y} v_z(t'') B_0 y(t'') dt''] dt', \qquad (10)$$

where $m$ is now the electron mass. The electron velocity is given by

$$v_x(t) = v_{x0} - \int_0^t \frac{q}{mL_x} v_z(t') B_{y0} x(t') dt', \qquad (11)$$

$$v_y(t) = v_{y0} + \int_0^t \frac{q}{mL_y} v_z(t') B_0 y(t') dt', \qquad (12)$$

$$v_z(t) = v_{z0} + \int_0^t \frac{q}{m} [E_0 + v_x(t') B_{y0} x(t')/L_x - v_y(t') B_0 y(t')/L_y] dt'. \qquad (13)$$

The initial and boundary conditions are $x(0)=x_0$, $x'(0)=v_{x0}$, $y(0)=y_0$, $y'(0)=v_{y0}$. Since the transit time of the electron in the small diffusion region is very short (Wagner et al., 1981), we can assume that during the transient time the change $\delta v_{z0}$ of $v_{z0}$ satisfies $\delta v_{z0} \ll v_{z0}$ or $v_z(t')$ is constant in Eqs. (9)-(12). As to be numerically verified in Section 4, the corresponding change in $v_z$ is even smaller. Eqs. (9)-(13) then yield

$$x(t) = x_0 \cosh(t/\tau_{dx}) + v_{x0} \tau_{dx} \sinh(t/\tau_{dx}), \qquad (14)$$

$$y(t) = y_0 \cos(t/\tau_{dy}) + v_{y0} \tau_{dy} \sin(t/\tau_{dy}), \qquad (15)$$

$$v_x(t) = x_0 \sinh(t/\tau_{dx})/\tau_{dx} + v_{x0} \cosh(t/\tau_{dx}), \qquad (16)$$

$$v_y(t) = -y_0 \sin(t/\tau_{dy})/\tau_{dy} + v_{y0} \cos(t/\tau_{dy}), \qquad (17)$$

$$\begin{aligned} v_z(t) = \; & v_{z0} + \frac{q}{m} E_0 t \\ & - \frac{1}{2v_{z0}} [(v_{x0}^2 + x_0^2/\tau_{dx}^2)\sinh^2(t/\tau_{dx}) + v_{x0} x_0 \sinh(2t/\tau_{dx})/\tau_{dx}] \\ & + \frac{1}{2v_{z0}} [(v_{y0}^2 - y_0^2/\tau_{dy}^2)\sin^2(t/\tau_{dy}) + v_{y0} y_0 \sin(2t/\tau_{dy})/\tau_{dy}], \end{aligned} \qquad (18)$$

where $\tau_{dx} = \sqrt{\dfrac{mL_x}{qv_{z0}B_{y0}}}$ and $\tau_{dy} = \sqrt{\dfrac{mL_y}{qv_{z0}B_0}}$ are characteristic times for charge particles on the x, y directions in the diffusion region, respectively. From Eqs. (14) and (15), they are indicated that the electron oscillates in the $y$ direction, but it is accelerated in the $x$ direction. If we assume $\tau_1$ to be the time when the electron leaves the box in the $x$ direction, from Eq. (14) and $x(\tau_1) = L_0$ we get

$$\tau_1 = \tau_{dx} \ln\left\{\left[L_0 + \sqrt{L_0^2 + v_{x0}^2 \tau_{dx}^2 - x_0^2}\right] / (v_{x0}\tau_{dx} + x_0)\right\}. \tag{19}$$

In order to see the acceleration process in more detail, we reasonably assume that thermal effects can be neglected. Thus, the initial in-plane velocity of an electron in the diffusion region is nearly zero. Considering the separation of the electron motion in the x and y directions, we only need to examine the electron motion in the x direction. The transit time then becomes

$$\tau = \tau_{dx} \ln \frac{L_0 + \sqrt{L_0^2 - x_0^2}}{x_0} . \tag{20}$$

Since $x_0$ can be anywhere inside the box, the averaged transit time is

$$\bar{\tau} = \int_0^{L_0} \tau dx_0 / L_0 = \frac{\pi}{2} \tau_{dx}, \tag{21}$$

so that the effective resistivity is given by

$$\eta_e = \frac{2}{\pi}\sqrt{\frac{mv_z B_{y0}}{q^3 n^2 L_x}} . \tag{22}$$

If the particle is farther away from the X line, $v_{x0}$ cannot be ignored, and the transit time is

$$\bar{\tau} = \frac{1}{L_0}\int_0^{L_0} \tau dx_0 = \tau_{dx}\left[\tan^{-1}\frac{L_0}{\delta} + \frac{\delta}{L_0}\ln\frac{(L_0+\delta)\delta}{L_0^2 + \delta^2}\right], \tag{23}$$

where $\delta = v_{x0}\tau_{dx}$. The corresponding effective resistivity is

$$\eta_{general} = \frac{m}{nq^2 \tau_{dx}}\frac{1}{\tan^{-1}\frac{L_0}{\delta} + \frac{\delta}{L_0}\ln\frac{(L_0+\delta)\delta}{L_0^2 + \delta^2}}. \tag{24}$$

Outside the region, $\delta$ is much larger (more precisely, $v_{x0}$ is much larger and $v_{z0}$ is smaller), making effective resistivity much smaller there. Since $v_{z0}$ is smaller outside the region, the assumption $\delta v_z \ll v_{z0}$ may breakdown. That is, our interaction model is applicable only in the small diffusion region around the X line.

If we include the ion motion, the current in Eq. (2) can be rewritten as

$$J_{z\delta} = nq^2 E_z(\tau_e)/m_e \quad \tau_i/m_i \ . \tag{25}$$

where we have assumed that the plasma is quasi-neutral. Eq. (3) then becomes

$$\eta_{tot} = \frac{m_e m_i}{nq^2} \frac{1}{m_i \tau_e + m_e \tau_i}. \tag{26}$$

Substituting Eq. (22), we get

$$\eta_{tot} = \eta_e \left(1 + \sqrt{\frac{m_e v_{ze}}{m_i v_{zi}}}\right)^{-1}, \tag{27}$$

where $v_{ze}$ and $v_{zi}$ are the local electron and ion velocities in the $z$ direction.

## 3 Simulation Model

We have performed 2.5D PIC simulations for electrons on the ($x,y$) plane by assuming $\partial/\partial z = 0$. For convenience, we use the charge-conservation scheme (CCS) (Villasenor & Buneman, 1992) instead of solving the Poisson equation, and the finite difference time domain method (FDTD) to solve the other Maxwell's equations. The particles are driven by the electric and Lorenz forces and the corresponding equations used in the PIC simulations are

$$\nabla \times \boldsymbol{E} = -\frac{\partial \boldsymbol{B}}{\partial t}, \tag{28}$$

$$\nabla \times \boldsymbol{B} = \varepsilon_0 \frac{\partial \boldsymbol{E}}{\partial t} + \mu_0 \boldsymbol{J}, \tag{29}$$

$$\frac{d\boldsymbol{p}_j}{dt} = q_j(\boldsymbol{E} + \boldsymbol{v}_j \times \boldsymbol{B}), \tag{30}$$

where $c$ is the light speed, $\boldsymbol{J} = n_i q_i \boldsymbol{V}_i + n_e q_e \boldsymbol{V}_e$, $\boldsymbol{V}_j$ ($j = i,e$) is the bulk velocity of species $j$, $\boldsymbol{v}_j$ and $\boldsymbol{p}_j = m_j \boldsymbol{v}_j$ are the particle velocity and momentum, respectively. The variables are normalized as follows: $x/d_{i0} \to x$, $(\boldsymbol{V}_j, \boldsymbol{v}_j)/v_{Ai0} \to (\boldsymbol{V}_j, \boldsymbol{v}_j)$, $\omega_{ci0} t \to t$, $\boldsymbol{B}/B_0 \to \boldsymbol{B}$, $\boldsymbol{E}/E_0 \to \boldsymbol{E}$, $\boldsymbol{J}/J_0 \to \boldsymbol{J}$, $n/n_0 \to n$, $\boldsymbol{p}_j/m_e v_{Ai0} \to \boldsymbol{p}_j$, where $d_{i0} = c/\omega_{pi0} = c/\sqrt{n_0 q_i^2/\mu_0 m_i}$, $v_{Ai0} = B_0/\sqrt{\mu_0 n_{i0} m_i}$, $\omega_{ci0} = q_i B_0/m_i$, $E_0 = v_{Ai0} B_0$, and $J_0 = n_0 q_0 v_{Ai0}$.

Our 2D simulation domain is $-D_x/2 \le x \le D_x/2$, $-D_y/2 \le y \le D_y/2$, where $D_x = 12.8 d_{i0}$, $D_y = 6.4 d_{i0}$, $dx = dy = 0.01 d_{i0}$. Closed boundary condition is adopted in the $y$ direction and periodic boundary condition is used in the $x$ direction. The time step is $\omega_{ci0}\Delta t = 0.0002$, and the duration of the simulations is $\omega_{ci0}t = 40$, corresponding to 200,000 time steps. Nearly 82 million particles for each species are used in this simulation. We also assume $v_{Ai0}/c = 0.05$ and $\beta = 0.2$. The ion-to-electron mass ratio $M_{ie} = m_i/m_e$ is from 25 to 400, and the ion-to-electron initial temperature ratio $T_{ie} = T_i/T_e = 5$.

We shall use the Harris equilibrium as the initial configuration. The initial magnetic field is given by

$$B_x = -B_0 \tanh(y/b_0), \quad B_y = B_z = 0, \tag{31}$$

and the initial density profile is

$$n = n_0/\cosh(y/b_0)^2 + n_b, \tag{32}$$

where $B_0 = 1.0$, $b_0 = 0.5$, $n_0 = 1.0$, $n_b = 0.2$, and $b_0$ is the width of the current sheet with the current intensity given by

$$I_z = B_0/b_0 \cosh(y/b_0)^2. \tag{33}$$

In order to shorten the initial stage in the simulation, we impose a small periodic excitation in the initial system, such that Eq. (31) and (33) become

$$B_x = -B_0 \tanh(y/b_0) - \varepsilon\pi\cos(2\pi x/D_x)\sin(\pi y/D_y)/D_y, \tag{34}$$

$$B_y = 2\varepsilon\pi\sin(2\pi x/D_x)\cos(\pi y/D_y)/D_x, \quad B_z = 0, \tag{35}$$

$$I_z = B_0/b_0 \cosh(y/b_0)^2 + \varepsilon\pi^2 \cos(2\pi x/D_x)\cos(\pi y/D_y)(1/D_y^2 + 4/D_x^2), \tag{36}$$

where $\varepsilon = 0.01$.

Pressure balance yields

$$P + \frac{B^2}{2} = (1+\beta)\frac{B_0^2}{2}, \tag{37}$$

where $P$ and $B$ are the local thermal pressure and magnetic field, $\beta = P/(B^2/2)$, and $P$ is normalized by $B_0^2/2\mu_0$.

## 4 Numerical Results and Comparison

First, we consider $M_{ie} = 25$, i.e., the same as that for the Geospace Environment Modeling (GEM) MR challenge (Pritchett, 2001). Figure 2 shows the evolution of the induction electric field and reconnected magnetic flux at the X line. We can see that MR occurs at $t = 20-32$, followed by a nonlinear stage of the process. Figure 3 shows the current $J_z$ and the magnetic field lines at different times. During the MR, the current sheet is compressed around the X line, and then separated into two parts.

The electric field in the $z$ direction as from the 2D electron fluid momentum equation is

$$E_z = -\frac{m_e}{e}\left[\frac{\partial V_{ez}}{\partial t} + \bm{V}_e \cdot \nabla V_{ez}\right] - \frac{1}{n_e e}\left(\frac{\partial \Pi_{exz}}{\partial x} + \frac{\partial \Pi_{eyz}}{\partial y}\right) - (\bm{V}_e \times \bm{B})_z, \tag{38}$$

where the pressure tensor is given by $\bm{\Pi}_e = m_e \int (\bm{v}-\bm{V})(\bm{v}-\bm{V}) f_e(v) dv$, where $f_e(V)$ is the electron velocity distribution function. Figure 4 shows the contribution of each term in Eq. (38) in the current sheet when MR occurs. We see that the sum of the off-diagonal pressure tensors leads to 80% of the induction electric field, similar to Pritchett's result in 2001.

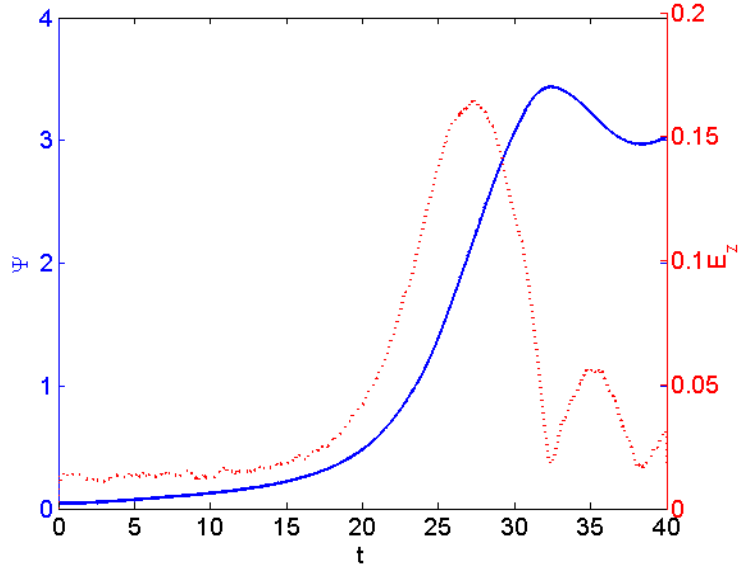

Figure 2　Evolution of reconnecting magnetic flux and the induction electric field on the X line. Here $\psi$ is normalized by $B_0 c / \omega_{pi0}$, and $E_z$ is normalized by $E_0$.

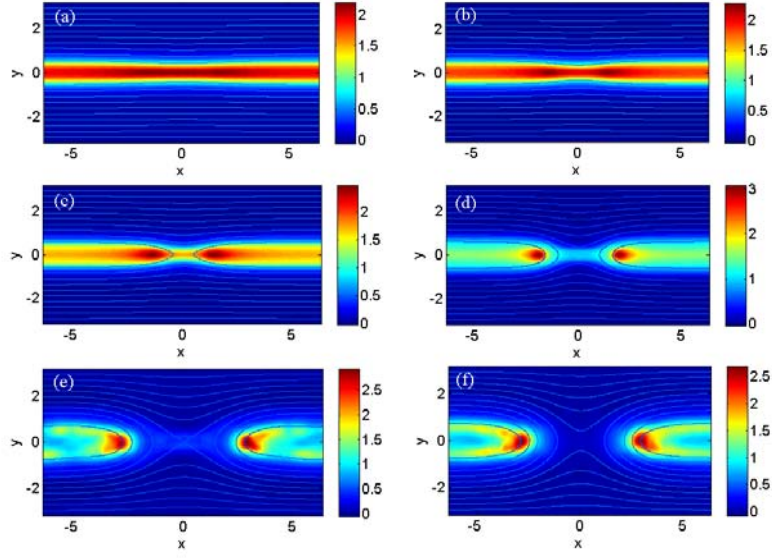

Figure 3    The distribution of the current density in the out of plane direction superposed with magnetic field lines at different simulation times. (a)-(f) shows the state of t=14, 18, 22, 26, 30, 34, respectively.

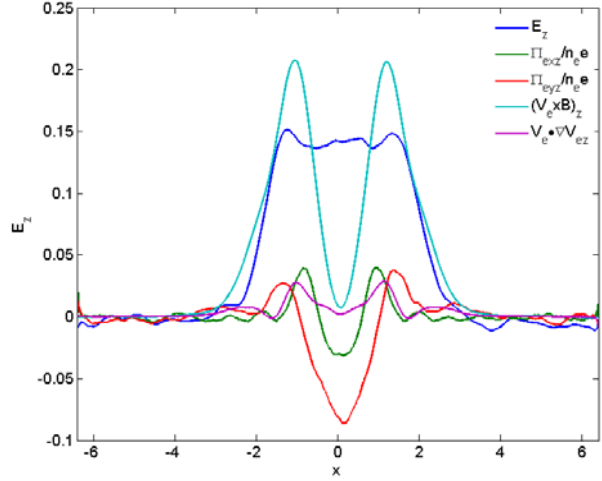

Figure 4    Contribution of each term from Eq. (38) in the current sheet along the x direction (at $y=0$) at the peak reconnection time $t=26$.

In order to verify the assumption $\delta v_z \ll v_{z0}$ used in Section 2, we compare the speeds of the particles which are just before entering and after leaving the electron diffusion region. Figure 5 shows the distribution of electron velocity variation $f(\delta v_z)$ during $t=28$ to 29 in the peaked

MR period. Here, $\delta v_z = v_{z1} - v_{z0}$, where $v_{z0}$ and $v_{z1}$ are the electron velocity when it is just before entering and after leaving the diffusion region, respectively. The mean of this distribution is 0.0470, and the variance is 0.2937. Electrons with $|\delta v_z/v_{z0}| \leq 0.2$ constitute 78.33% of the total ejected electrons, implying that most of the electrons suffer little change in the $z$ component direction velocity. In the earlier MR stage, such as from $t = 18$ to 19, the percent of electrons with $|\delta v_z/v_{z0}| \leq 0.2$ is 90.96%. Thus, the velocity changes $\delta v_z$ for the majority of electrons are limited when they stay in the smaller diffusion region, so that our assumption in the derivation of the effective resistivity is justified. It is clearly also valid for ions, whose velocities are much less.

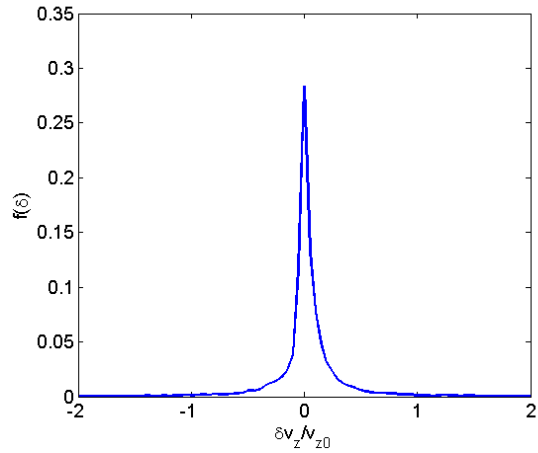

Figure 5　Distribution of $\delta v_z$ during $t =28$ to 29.

Figure 6 shows the time evolution of the average energy per electron for different components. "entering" and "leaving" means for electrons just before entering and after leaving the electron diffusion region, respectively. We can see that the difference of the average energy per electron in the in-plane component for the "entering" and "leaving" electrons is relatively small at all times, which agrees with our assumption that the in-plane electric field is nearly zero in Eq. (5). The energy gain of electrons in the z component increases with development of MR during the period in the diffusion region. The energy gain is about 20% when MR reaches its peak, which means the net change of the velocity in the z direction is about 10%. Therefore, it is further confirmed that our assumption $\delta v_z \ll v_{z0}$ is valid. The energy gain of electrons disappears after the fast reconnection stage ends. This behavior can be attributed to the fact that in the period of FMR, the induced electric field in the $z$ direction is strong around the X line. On the other hand, the magnetic field is weak in this region and they are not sufficient to alter the trajectory of the hot electrons, which leads to electrons continuously accelerated in the z direction.

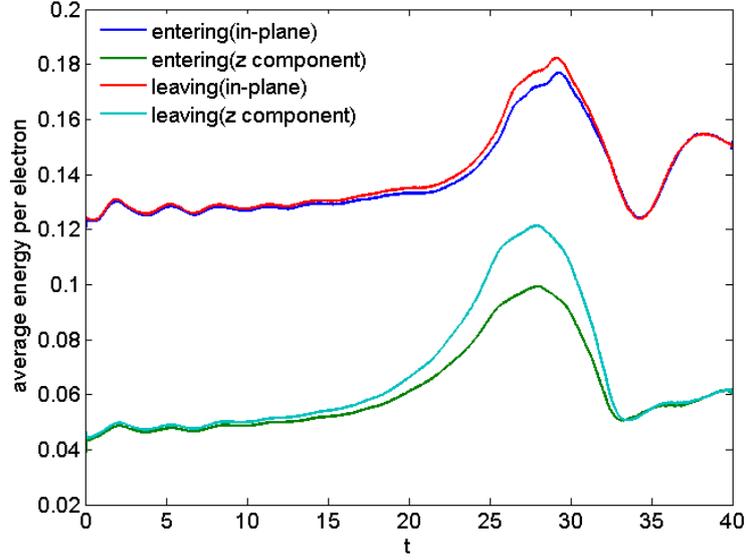

Figure 6  Time evolution of the average energy per electron for different components. "entering" and "leaving" means for electrons just before entering and after leaving the diffusion region, respectively.

Figure 7 shows time evolutions of the effective resistivity in the electron diffusion region for different mass ratios $M_{ie}$. Since larger $M_{ie}$ corresponds to a longer linear growth phase, we use larger initial excitation ($\varepsilon = 0.05$) for $M_{ie} = 256$ and 400 to shorten simulation time. The other parameters remain the same. The theoretical $\eta_e$ and $\eta_{tot}$, as well as the simulation result $\eta_s = E_z / J_z$, are all normalized by $B_0 / (n_0 q_0)$.

We can see that as MR enters into the fast reconnection phase, the effective resistivity exhibits quickly enhancement and the tendencies are almost the same for all the three effective resistivities. With increasing $M_{ie}$, not only do the peak values of the effective resistivity decrease, but also the difference between $\eta_e$ and $\eta_{tot}$ decreases, as can also be seen in Eq. (27). Since the PIC simulation involves larger noise level compared to the MHD simulation, the resistivity $\eta_s$ directly from simulation fitting the modeled effective resistivity are reasonably well.

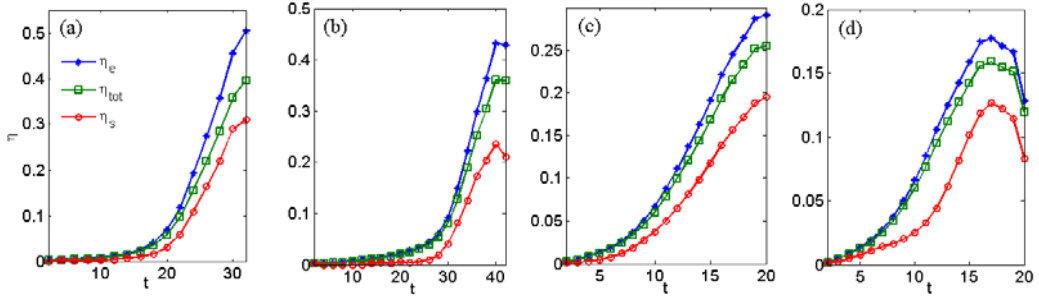

Figure 7 Time evolutions of the effective resistivity from the model and PIC simulation for different $M_{ie}$'s. From left to right, (a)-(d) represents the case when $M_{ie}$ equals to 25, 100, 256, 400, respectively. The time interval between the points is $dt=2$ in (a)-(b), and $dt=1$ in (c)-(d).

We also present the electric field $E_z$ and the current density $J_z$ in the z direction directly from the simulation in Figure 8. It is found that the changes of $E_z$ and $J_z$ are not in phase with the effective resistivity $\eta_s$. The effective resistivity further increases after the reconnection electric field $E_z$ decreases. This is because the current density $J_z$ always decreases and the decreasing speed is proportional to the electric field $E_z$, which is mainly attributed to the decrease of the electron density in the diffusion region.

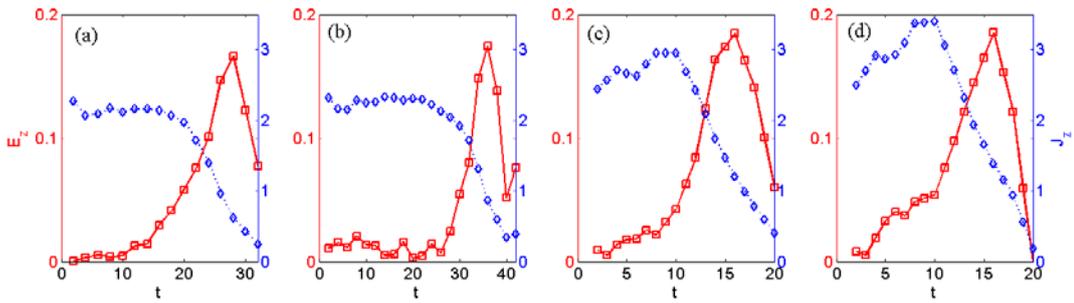

Figure 8 Time evolutions of the electric field and the current density in the z direction for different $M_{ie}$'s. From left to right, (a)-(d) represents the case when $M_{ie}$ equals to 25, 100, 256, 400, respectively. The time interval between the points is $dt=2$ in (a) and (b), the others is $dt=1$.

The inertial conductivity $\sigma_i$ from Eq. (14) of Speiser (1970) is

$$\sigma_i = \frac{ne^2}{m}\tau = \frac{ne^2}{m}\frac{L}{v} = \left(\frac{Ln^2e^3}{2mE}\right)^{1/2}, \quad (39)$$

where $L$ is the length of the accelerating region, $v$ is the particle velocity and $E$ is the electric field. Here $m$ we take as the electron mass. It is evident that this model is not able to implement into MHD because we first have to know the resistivity to calculate the electric field. Another problem is that the estimated resistivity from this model is about 5 times larger than the numerical results from $\eta_s = E_z / J_z$.

## 5 Summary

This paper introduces a simple model for energy conversion in FMR. Using the simple equation $E = \eta J$, we define a space-time dependent effective resistivity $\eta$ that can be obtained from analytical solutions of test electron trajectories in the diffusion region. We find that $\eta$ rises with development of MR, but still increases after MR reaches its peak. It then falls and finally reaches a low value. The results from the model agree fairly well with that from the PIC simulations. It is also found that the difference of effective resistivities from the PIC simulation and our model tends to be smaller with the increase of the ratio of ion and electron.

We wish this paper can give a new view on anomalous resistivity in MHD simulation, whose idea is derived from a collisionless fast reconnection model, and the physical meaning is reasonable.

## Acknowledgements

We would like to thank Prof. M. Y. Yu for his useful advices and polishing on this paper, and we would also like to thank Prof. L. C. Lee and H. W. Zhang for their useful suggestions. This work is supported by the National Natural Science Foundation of China under Grant No. 41474123, National Magnetic Confinement Fusion Science Program of China under Grant No. 2013GB104004 and 2013GB111004, the Special Project on High-performance Computing under the National Key R&D Program of China No. 2016YFB0200603, Fundamental Research Fund for Chinese Central Universities. The simulation data is available upon request.

# References

Bhattacharjee, A. (2004), Impulsive magnetic reconnection in the earth's magnetotail and the solar corona, *Annual review of astronomy & astrophysics*, 42(1), 365-384. doi:10.1146/annurev.astro.42.053102.134039

Birn, J., & Hesse, M. (1991), The substorm current wedge and field-aligned currents in MHD simulations of magnetotail reconnection, *Journal of geophysical research*, 96(A2), 1611-1618. doi:10.1029/90JA01762

Cai, H. J., & Lee, L. C. (1997), The generalized Ohm's law in collisionless magnetic reconnection, *Physics of plasmas*, 4(3), 509-520. doi:10.1063/1.872178

Deng, X. H., & Matsumoto, H. (2001), Rapid magnetic reconnection in the earth's magnetosphere mediated by whistler waves, *Nature*, 410(6828), 557-60. doi:10.1038/35069018

Dungey, J. W. (1961), Interplanetary magnetic field and the auroral zones, *Physical Review Letters*, 6(2), 47-48. doi:10.1103/PhysRevLett.6.47

Furth, H. P., Rutherford, P. H., & Selberg H. (1973), Tearing mode in the cylindrical tokamak, *Physics of fluids*, 16(16), 1054-1063. doi:10.1063/1.1694467

Goldstein, M. L., Matthaeus, W. H., & Ambrosiano, J. J. (1986), Acceleration of charged particles in magnetic reconnection: solar flares, the magnetosphere, and solar wind, *Geophysical research letters*, 13(3), 205-208. doi:10.1029/GL013i003p00205

Hesse, M., Forbes, T. G., & Birn, J. (2005), On the relation between reconnection magnetic flux and parallel electric fields in the solar corona, *The Astrophysical Journal*, 631(2), 1227-1238. doi:10.1086/432677

Hsieh, M. H., Tsai, C. L., Ma, Z. W., & Lee, L. C. (2009), Formation of fast shocks by magnetic reconnection in the solar corona, *Physics of plasma*, 16(9), 092901-092901-9. doi:10.1063/1.3212889

Kopp, R. A., & Pneuman, G. W. (1976), Magnetic reconnection in the corona and the loop prominence phenomenon, *Solar Physics*, 50(1), 85-98. doi:10.1007/BF00206193

Malyshkin, L. M., Linde, T., & Kulsrud, R. M. (2005), Magnetic reconnection with anomalous resistivity in two-and-a-half dimensions. I. Quasistationary case, *Plasma of physics*, 12(10), 123. doi:10.1063/1.2084847

Moses, R. W., Finn, J. M., & Ling, K. M. (1993), Plasma heating by collisionless magnetic reconnection: analysis and computation, *Journal of geophysical research*, 98(A3), 4013-4040. doi:10.1029/92JA02267

Oieroset, M., Phan, T. D., Fujimoto, M., Lin, R. P., & Lepping, R. P. (2001), In situ detection of collisionless reconnection in the earth's magnetotail, *Nature*, 412(6845), 414. doi:10.1038/35086520

Pritchett, P. L. (2001), Geospace environment modeling magnetic reconnection challenge: simulations with a full particle electromagnetic code, *Journal of geophysical research*, 106(A3), 3783-3798. doi:10.1029/1999JA001006

Speiser, T. W. (1970), Conductivity without collisions or noise, *Planetary &Space Science*, 18(4), 613-622. doi:10.1016/0032-0633(70)90136-4

Ugai, M. (1984), Self-consistent development of fast magnetic reconnection with anomalous plasma resistivity, *Plasma physics and controlled fusion*, 26(12B), 1549. doi:10.1088/0741-3335/26/12B/010

Vasyliunas, V. M. (1975), Theoretical models of magnetic field line merging, *Reviews of Geophysics*, 13(1), 303-336. doi:10.1029/RG013i001p00303

Villasenor, J., & Buneman, O. (1992), Rigorous charge conservation for local electromagnetic field solvers, *Computer Physics Communications*, 69(2), 306-316. doi:10.1016/0010-4655(92)90169-Y

Wagner, J. S., Gray, P. C., Kan, J. R., Tajima T., & Askasofu, S. I. (1981), Particle dynamics in reconnection field configurations, *Planetary &Space Science*, 29(4), 391-397. doi:10.1016/0032-0633(81)90082-9

Wang, S., & Ma, Z. W. (2015), Influence of toroidal rotation on resistive tearing modes in tokamaks, *Physics of plasmas*, 22(12), 2251-S202. doi:10.1063/1.4936977